\documentclass[english,reprint]{revtex4-1}
\usepackage[latin9]{inputenc}
\usepackage{amsmath}
\usepackage{amssymb}
\usepackage{graphicx}
\usepackage{esint}

\makeatletter
 
 \@ifundefined{textcolor}{}
 {%
   \definecolor{BLACK}{gray}{0}
   \definecolor{WHITE}{gray}{1}
   \definecolor{RED}{rgb}{1,0,0}
   \definecolor{GREEN}{rgb}{0,1,0}
   \definecolor{BLUE}{rgb}{0,0,1}
   \definecolor{CYAN}{cmyk}{1,0,0,0}
   \definecolor{MAGENTA}{cmyk}{0,1,0,0}
   \definecolor{YELLOW}{cmyk}{0,0,1,0}
 }

\usepackage{babel}

\usepackage{babel}

\makeatother

\usepackage{babel}
\begin{document}

\title{Memory effect induced macroscopic-microscopic entanglement}

\author{Qingxia Mu$^{1,2}$}

\email{qingxiamu@ncepu.edu.cn}

\author{Xinyu Zhao$^{2}$}

\email{xyzacademic@gmail.com}

\author{Ting Yu$^{2,3}$}

\email{Ting.Yu@stevens.edu}

\affiliation{$^{1}$Mathematics and Physics Department, North China Electric Power
University, Beijing 102206, China}

\affiliation{$^{2}$Center for Controlled Quantum Systems and the Department of
Physics and Engineering Physics, Stevens Institute of Technology,
Hoboken, New Jersey 07030, USA}

\affiliation{$^{3}$Beijing Computational Science Research Center, Beijing 100094, China}

\begin{abstract}
We study optomechanical entanglement between an optical cavity field
and a movable mirror coupled to a non-Markovian environment.  The non-Markovian 
quantum state diffusion (NMQSD) approach and the non-Markovian master equation  
are shown to be useful in investigating the entanglement generation between the 
cavity field and the movable mirror. The simple model presented in this paper
demonstrates several interesting properties of optomechanical
entanglement that are associated with environment memory effects. It is evident
that the effective environment central frequency can be used to modulate 
the optomechanical entanglement. In addition, we show that the maximum entanglement
may be achieved by properly choosing the effective detuning
which is significantly dependent on the strength of the memory effect of the
environment.
\end{abstract}
\maketitle

\section{Introduction}

Macroscopic quantum coherence has a long history that may date back
to the famous Schr\"{o}dinger's cat paradox \cite{Cat}. Although current
research in quantum mechanics does not impose a strict boundary between
quantum and classical realms, realising reliable microscopic-macroscopic entanglement
is still a challenge due to the so-called decoherence processes which
are especially severe for a macroscopic object. This explains that entanglement
is most commonly observed in the microscopic world. In recent years,
several attempts in establishing entanglement in the macroscopic or
mesoscopic systems have been made \cite{Macro1,Macro2,Macro3,Macro4,Macro5,Macro6,Macro7,Macro8,Macro9,Macro10,Chou2008PRE,Chou2008Physica}.
In the same spirit, quantum entanglement between a microscopic object
and a macroscopic object is expected to be a useful resource
for the emerging quantum technology such as quantum information processing
and quantum computing. In addition, a deeper understanding of the micro-macro entanglement 
and its decoherence process may be important for a better understanding of transition from
classical to quantum realms \cite{QtoC,QtoC2}.  Apart from the motivation from the theoretical
research activities, the latest developments in experimental entanglement generation, control and
manipulation have provided a direct impetus for further explorations
of this important setting based on optomechanical systems \cite{Experi1,OptoCool1,Experi3,Opto1,Opto2,Opto3}.

The radiation pressure in an optical cavity
is capable of producing entanglement between the quantized cavity modes
(microscopic system) and a movable mirror \cite{Opto4}. When the mass of the mirror 
is in the macroscopic scale,  such an optomichanical system 
provides a natural testing bed for macroscopic quantum 
mechanical phenomena \cite{ChenJPBreview,Opto5,Opto6,Opto7,Opto8,Opto9,Opto10}.
Theoretically, Markov Langevin equations or the corresponding master equations 
may be used to deal with an optomechanical system when the environmental
noises can be treated as a weak perturbation and the noisy memory
effect can be ignored \cite{Markov}. 
However, it becomes clear that,  from both theoretical and
experimental viewpoint, the memory effects plays a pivotal role in
micro-macro entanglement dynamics
in non-Markovian regimes \cite{Chou2008PRE,Paz2008,NMOpto}. Notably, 
a recent experimental observation \cite{Groblacher_2015_NMexp} has explicitly
shown that the heat bath coupled to the optomechanical system is in non-Markovian regimes. Moreover, the non-Markovian properties are shown to be useful in preserving optomechanical entanglement \cite{Cheng_2016_NMOpt}.
In addition, an environmental engineering technique for a non-Markovian bath demonstrated 
in an optical experiment \cite{Liu_2011_NatPhys} has suggested a promising future in manipulating non-Markovian
environments to control quantum dynamics of the system of interest. Hence, it is highly desirable to develop a systematic 
approach to investigate the dynamics of optomechanical system in non-Markovian
regime. 

The purpose of this paper is to investigate the entanglement between
the light field in a Fabry-P\'{e}rot cavity and one movable reflection
mirror of the cavity (Fig.~1). The movable mirror is assumed to be
embedded in a non-Markovian environment modeled by a bosonic 
bath. We shall begin our discussion with an exact quantum description
of the optomechanical system consisting of cavity modes and the movable
mirror. The advantage of the exact treatment is that the memory
effect in this model can be treated in a systematic way without
introducing any {\t{a}d hoc} parameters to represent the environmental
noises. We shall use the non-Markovian quantum state diffusion (NMQSD)
equation to solve quantum open systems coupled to a non-Markovian
bosonic or fermionic 
environment \cite{Yang2012,QSD98,QSD99,Yu1999,QBM,Yu_FiniteT,Jing-Yu2010,Xinyu2011,ZhaoFB,ShiFB,ChenFB,Ncav,chen2014,Xu2014}.
Such a stochastic approach provides a very powerful tool in both analytical
treatments and numerical simulations, especially in dealing with the non-Markovian
perturbation and solving the corresponding master equation for the
open quantum systems. With our approach, the model considered in this
paper can be solved efficiently to exhibit the non-Markovian properties
that affect the dynamics of optomechanical entanglement. More specifically,
our results show that the environmental memory can significantly alter
the speed of the optomechanical entanglement generation between the
cavity field and the movable mirror. Our approach can also incorporate
importantly the high frequency back reaction of the environment, which
will be shown to significantly preserve the generated entanglement.
Furthermore, we show that a proper choice of effective detuning is
also crucial for optimizing the optomechanical entanglement.

The paper is organized as follows. In Sec.~\ref{sec:Model-and-Solution}, we present the interacting model and derive the master equation based on the NMQSD equation. Sec.~\ref{sec:Numerical-results-and} analyzes in detail the environmental memory effects on the entanglement between the mechanical mode and the intracavity mode. In particular, we show how the effective environment central frequency can be used to modulate 
the optomechanical entanglement. In addition, we show that the maximum entanglement
may be achieved by properly choosing the effective detuning which is significantly dependent on the strength of the memory effect of the
environment. Finally, we conclude the paper in Sec.~\ref{conclusion}. Some details about the equations of motions for the mean values 
are left to the the Appendixes.

\section{\label{sec:Model-and-Solution}Model and Solution}

\begin{figure}
\noindent \includegraphics[clip,width=1\columnwidth]{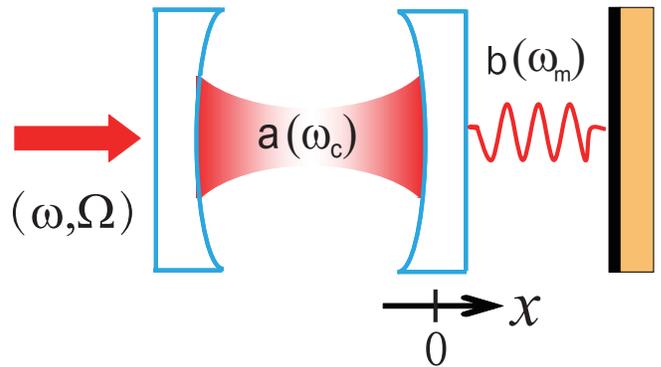} \caption{\label{fig:setup}(color online) Schematic diagram of a typical optomechanical
system, in which an optical cavity driven by a coherent laser is coupled
to a mechanical mode.}
\end{figure}

We consider a single-sided optomechanical system, with a mechanical
mode coupled to an optical mode which is driven by a coherent laser,
as shown schematically in Fig.~\ref{fig:setup}. The Hamiltonian of this system
may be written as $H_{1}=\omega_{c}a_{1}^{\dagger}a_{1}+\omega_{m}b_{1}^{\dagger}b_{1}+ga_{1}^{\dagger}a_{1}(b_{1}+b_{1}^{\dagger})+\Omega_{d}(a_{1}e^{i\omega t}+a_{1}^{\dagger}e^{-i\omega t})$
\cite{Law1994}, where $a_{1}$ and $b_{1}$ are annihilation operators
of the cavity field and mechanical mode, with respective resonant
frequencies $\omega_{c}$ and $\omega_{m}$. The parameter $g$ is
the single-photon optomechanical coupling strength, and $\Omega_{d}$
is the driving rate of the coherent laser with frequency $\omega$.
We assume that the intracavity field is strong enough that the Hamiltonian
can be linearized with $a_{1}\equiv a+\alpha$, $b_{1}\equiv b+\beta$.
Here $a$ and $b$ represent quantum fluctuations of optical and mechanical
modes around their mean values $\alpha$ and $\beta$, respectively.
They are determined by $\left[i(\omega-\omega_{c})-ig(\beta+\beta^{*})-\kappa_{a}\right]\alpha-i\Omega_{d}=0$
and $-i\omega_{m}\beta-ig|\alpha|^{2}=0$, where $\kappa_{a}$ is
classical leakage rate of the cavity. The Hamiltonian of the system
can be linearized as \cite{Opto2} 

\begin{equation}
H_{S}=-\Delta a^{\dagger}a+\omega_{m}b^{\dagger}b+G(a^{\dagger}+a)(b^{\dagger}+b),\label{eq:Hs}
\end{equation}
where $G=\alpha g$ is the effective coupling rate, $\Delta=\omega-\omega_{c}+2G^{2}/\omega_{m}$
is the optomechanical-coupling modified detuning.

We assume that the optomechanical system is coupled to a bosonic bath
which can be described by a set of harmonic oscillators as 
\begin{equation}
H_{B}=\sum_{j}\omega_{j}c_{j}^{\dagger}c_{j},\label{eq:Hb}
\end{equation}
where $c_{j}$ and $c_{j}^{\dagger}$ are annihilation and creation
operators satisfying $[c_{j},c_{j'}^{\dagger}]=\delta_{j,j'}$. The
interaction between the system and the bosonic bath is given by 
\begin{equation}
H_{I}=\sum_{j}g_{j}(Lc_{j}^{\dagger}+L^{\dagger}c_{j}),\label{eq:HI}
\end{equation}
where $g_{j}$ are the system-bath coupling strength, and $L=b$ describing
the damping of the mirror. Here, the interaction is written in a rotating
wave approximation (RWA) form, a more general interaction for mirror
and bath should be $H_{I}^{\prime}=\sum_{j}g_{j}(b+b^{\dagger})(c_{j}^{\dagger}+c_{j})$.
If the coupling strength is weak comparing to the system ($g_{j}\ll\omega_{m}$)
\cite{Joshi2014_RWA,Cresser1992}, this approximation is valid. Actually
the interaction $H_{I}^{\prime}$ can be also incorporated in the
NMQSD approach. The method of solving $H_{I}^{\prime}$ type of interaction
can be found in Ref. \cite{QBM}.

More general discussions on the issue may include the decoherence channels concerning the cavity leakage 
and the thermal damping of the mirror. For the sake of reducing technical complexity of our model, we will
exclusively be focused on the vacuum environment, leaving more complete model description to the 
appendix \ref{sec:App1} where we provide a full solution of the NMQSD equation. Our major concern 
in this paper is not the temperature effect on the decoherence rate, but rather on the non-Markovian properties of the environment. 

Assuming that the system and the environment are initially uncorrelated,
it can be proved that the state of the optomechanical system can be
represented by a stochastic pure state called quantum trajectory,
$|\psi_{t}(z^{*})\rangle$, governed by the NMQSD equation \cite{QSD98,QSD99}
\begin{equation}
\partial_{t}|\psi_{t}(z^{*})\rangle=\left[-iH_{S}+bz_{t}^{*}-b^{\dagger}\int_{0}^{t}ds\alpha(t,s)\frac{\delta}{\delta z_{s}^{*}}\right]|\psi_{t}(z^{*})\rangle,\label{eq:QSD}
\end{equation}
where $\alpha(t,s)=\sum_{j}|g_{j}|^{2}e^{-i\omega_{j}(t-s)}$ is the
environmental correlation function, and $z_{t}^{*}=-i\sum_{j}g_{j}z_{j}^{*}e^{i\omega_{j}t}$
is a complex Gaussian process satisfying $M[z_{t}]=M[z_{t}z_{s}]=0$
and $M[z_{t}^{*}z_{s}]=\alpha(t,s)$. Here $M[\cdot]\equiv\int\frac{dz^{2}}{\pi}e^{-|z|^{2}}[\cdot]$
stands for the statistical average over the noise $z_{t}$. Note that
the above dynamical equation (\ref{eq:QSD}) contains a time-nonlocal
term depending on the whole evolution history from $0$ to $t$. For
the purpose of practical applications, we can replace the functional
derivative contained in Eq.~(\ref{eq:QSD}) with a time-dependent
operator $O$ satisfying $\frac{\delta|\psi_{t}(z^{*})\rangle}{\delta z_{s}^{*}}=O(t,s,z^{*})|\psi_{t}(z^{*})\rangle$.
Then the NMQSD equation can be transformed to 
\begin{equation}
\partial_{t}|\psi_{t}(z^{*})\rangle=\left[-iH_{S}+bz_{t}^{*}-b^{\dagger}\bar{O}(t,z^{*})\right]|\psi_{t}(z^{*})\rangle,\label{eq:QSD2}
\end{equation}
where $\bar{O}(t,z^{*})\equiv\int_{0}^{t}ds\alpha(t,s)O(t,s,z^{*})$
and the initial condition $O(t,s=t,z^{*})=b$ is satisfied.

It should be noted that Eq.~(\ref{eq:QSD2}) {[}as well as Eq. (\ref{eq:QSD}){]}
is derived directly from the microscopic Hamiltonian (\ref{eq:Hs}),
(\ref{eq:Hb}), and (\ref{eq:HI}) without any approximation. It is
the exact dynamic equation governing the dynamics of the optomechanical
system coupled to the environment, no matter the environment is in
Markov or non-Markovian regime. The environmental impact on the dynamics
of the optomechanical system is reflected on the terms $bz_{t}^{*}$
and $-b^{\dagger}\bar{O}(t,z^{*})$ in Eq. (\ref{eq:QSD2}). If these
two terms are zero, the equation is reduced to $\partial_{t}|\psi_{t}(z^{*})\rangle=-iH_{S}|\psi_{t}(z^{*})\rangle$,
which is the Schr\"{o}dinger equation for the closed system. Moreover,
the non-Markovian properties are reflected by the operators $\bar{O}$
in Eq. (\ref{eq:QSD2}). If there is no correlation between two separate
time points $t$ and $s$, namely $\alpha(t,s)=\delta(t,s)$, the
operator $\bar{O}$ is reduced to $\bar{O}=b$. As a result, Eq.~(\ref{eq:QSD2}) is reduced to the commonly used Markov quantum trajectory
equation \cite{Dalibardetal,Gisin-Percival}. Here, and throughout
the paper, the correlation function of the environment is chosen as
the Ornstein-Uhlenbeck (O-U) correlation function 
\begin{equation}
\alpha(t,s)=\frac{\Gamma\gamma}{2}e^{-(\gamma+i\Omega)|t-s|},
\end{equation}
in which the parameter $1/\gamma$ measures the memory time, $\Gamma$ is the environmental
decay rate, and $\Omega$
is the central frequency of the environment. The O-U type correlation function corresponds to the
Lorentzian spectrum density $J(\omega)=\frac{\Gamma\gamma^{2}/2\pi}{(\omega-\Omega)^{2}+\gamma^{2}}$
of the environment, which has been widely used in the research on
cavity optomechanics \cite{Opto2,Naeini_2013_Lorentzian1}. 
A more generic correlation function or spectrum
density may be needed in many other interesting situations. 
We would like to emphasize that our derivation is independent of a specific
form of the correlation functions $\alpha(t,s)$, so that Eq.~(\ref{eq:QSD2}) is applicable to an arbitrary correlation function.
The reason we use the O-U correlation function 
here is that it is convenient to observe the crossover properties of the non-Markovian and Markov transition 
by modulating the single parameter $\gamma$. If the memory time $1/\gamma$ is very small, $\alpha(t,s)$
is approximately reduced to $\alpha(t,s)\approx\delta(t,s)$, which
means the environment is reduced to a Markov environment.

The key to solving the dynamic equation (\ref{eq:QSD2}) is to find the operator
$O$.  The exact solution of $O$ operator contains all the non-Markovian
information for the environment. The exact $O$ is also essential for the derivation
of the corresponding exact master equation. According to Refs.~\cite{QSD98,QSD99},
$O$ satisfies  the following equation,
\begin{equation}
\frac{\partial}{\partial t}O=[-iH_{S}+Lz_{t}^{*}-L^{\dagger}\bar{O},O]-L^{\dagger}\frac{\delta}{\delta z_{s}^{*}}\bar{O}.\label{eq:dtO}
\end{equation}
Clearly, finding the exact $O$ operator for a particular model is not easy. Therefore, for most practical problems,  of central importance
 of applications is the perturbation approach \cite{Yu1999}. Notably, it is shown that the exact $O$ operator for the model in
this paper can be found,
\begin{equation}
O(t,s,z^{*})=\sum_{j=1}^{4}f_{j}(t,s)O_{j}+i\int_{0}^{t}ds'f_{5}(t,s,s')z_{s'}^{*}O_{5},\label{eq:O}
\end{equation}
where the basis operators are given by 
\begin{equation}
O_{1}=b,O_{2}=b^{\dagger},O_{3}=a,O_{4}=a^{\dagger},O_{5}=I,
\end{equation}
and $f_{j}\;(j=1\cdots5)$ are time-dependent coefficients. Substituting
Eq.~(\ref{eq:O}) to Eq.~(\ref{eq:dtO}), the differential
equations for the coefficients in the $O$ operator can be determined
as 
\begin{eqnarray}
\frac{\partial}{\partial t}f_{1}(t,s) & = & i\omega_{m}f_{1}+iGf_{3}-iGf_{4}+f_{1}F_{1},\nonumber \\
\frac{\partial}{\partial t}f_{2}(t,s) & = & -i\omega_{m}f_{2}+iGf_{3}-iGf_{4}-f_{2}F_{1}\nonumber \\
 &  & +2f_{1}F_{2}-f_{4}F_{3}+f_{3}F_{4}-F_{5}^{\prime},\nonumber \\
\frac{\partial}{\partial t}f_{3}(t,s) & = & -i\Delta f_{3}+iGf_{1}-iGf_{2}+f_{1}F_{3},\nonumber \\
\frac{\partial}{\partial t}f_{4}(t,s) & = & i\Delta f_{4}+iGf_{1}-iGf_{2}+f_{1}F_{4},\nonumber \\
\frac{\partial}{\partial t}f_{5}(t,s,s') & = & f_{1}F_{5}^{\prime}(t,s'),\label{eq:fi}
\end{eqnarray}
where $F_{j}(t)=\int_{0}^{t}ds\alpha(t,s)f_{j}(t,s)\;(j=1\cdots4)$ and
$F_{5}^{\prime}(t,s')=\int_{0}^{t}ds\alpha(t,s)f_{5}(t,s,s')$. The
boundary conditions are give by 
\begin{eqnarray}
 &  & f_{1}(t,s=t)=1,\nonumber \\
 &  & f_{2}(t,s=t)=f_{3}(t,s=t)=f_{4}(t,s=t)=0,\nonumber \\
 &  & f_{5}(t,s=t,s')=0,\nonumber \\
 &  & f_{5}(t,s,s'=t)=f_{2}(t,s).\label{eq:BC}
\end{eqnarray}

\begin{figure}
\noindent \includegraphics[clip,width=1\columnwidth]{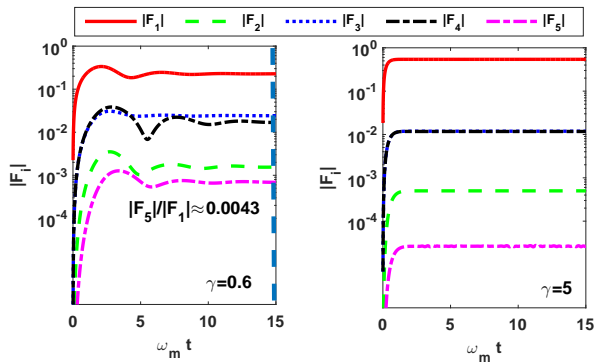} \caption{\label{fig:Fi}(color online) Time evolution of the coefficients in
$O$ operator. For the case $\gamma=0.6$, $\frac{|F_{5}|}{|F_{1}|}\approx0.43\%$
at $\omega_{m}t=15$.}
\end{figure}

In Eq.~(\ref{eq:QSD2}), the non-Markovian properties are reflected
by the correlation function. If the correlation is $\alpha(t,s)=\delta(t,s)$,
$\bar{O}$ is reduced to $\bar{O}=O_{1}=b$, as a result, Eq.~(\ref{eq:QSD2}) is reduced to the Markov quantum trajectory equation
investigated in Refs.~\cite{Dalibardetal,Gisin-Percival}. Clearly,
the additional terms $O_{i}(i=2,3,4,5)$ contribute to the non-Markovian
corrections. In Fig. \ref{fig:Fi}, we plot the time evolution of
the time-dependent coefficients $F_{i}$ ($i=1,2,3,4$) and $F_{5}(t)=\int_{0}^{t}ds^{\prime}\alpha(t,s^{\prime})F_{5}^{\prime}(t,s^{\prime})$.
From the right panel of Fig.~\ref{fig:Fi}, when $\gamma$
is increased, the environment is approaching the well-known Markov limit, hence, the non-Markovian
corrections $F_{i}\;(i=2,3,4,5)$ are becoming ignorable.  On the contrary, in the case
of small $\gamma$ as plotted in the left panel of Fig.~\ref{fig:Fi},
the non-Markovian corrections $F_{i}\;(i=3,4)$ become more notable
compared with the right panel. Fig.~\ref{fig:Fi} roughly shows the corrections
to the Markov case caused by the finite memory time and the transition
from non-Markovian to Markov regimes. Clearly, the non-Markovian environment
not only causes the additional terms $O_{i}(i=2,3,4,5)$, but also changes
the dynamical behavior especially in the early stage of the evolution.
Compared with the right panel,  we see that the left panel exhibits some transient
oscillation in early stage. This oscillatory evolution directly represents
the information exchange between the system and its environment due
to the memory effect. In the Markov limit,  the environment typically make the
system converge to a steady state quickly. As an interesting observation, our 
discussions later in the paper show that these oscillations eventually result in different 
entanglement generations. It is also notable that in both panels of Fig.~\ref{fig:Fi},
the fifth term $F_{5}$ always gives the smallest correction. Hence, this term
might be dropped in an approximation approach.

For the purpose of numerical simulation, one can directly simulate
the NMQSD equation (\ref{eq:QSD2}) together with the $O$ operator
given in equation (\ref{eq:O}). Repeatedly solving equation (\ref{eq:QSD2})
with stochastic noise $z_{t}^{*}$ and taking the statistical mean
of all the generated trajectories, the reduced density matrix can be recovered
as 
\begin{equation}
\rho_{t}=M[|\psi_{t}(z^{*})\rangle\langle\psi_{t}(z^{*})|].\label{eq:rdm}
\end{equation}
The advantage of using this pure state stochastic trajectories approach
is the required computational resource is reduced from $N^{2}$ 
(to storage density matrix) to $N$  (to storage pure state vector).
Alternatively, one can also use the NMQSD equation to derive the corresponding
exact master equation for the system by following the method in Ref.~\cite{QBM,chen2014}.
In this paper, we will take a straightforward step to truncate the
$O$ operator to the noise free terms, called the zeroth order approximation,
which turns out to be appropriate for many practical purposes as discussed
in Ref.~\cite{Xu2014}. Here, as shown before, the fifth term is typically much
smaller than the other four terms, so we take the first four terms of the
$O$ operator as an approximate $O$ operator 
\begin{equation}
\label{ofree}
O(t,s,z^{*})\approx O^{(0)}(t,s)=\sum_{j=1}^{4}f_{j}(t,s)O_{j}.
\end{equation}
More systematic discussions on the validity of this approximation has been 
discussed in Ref.~\cite{Xu2014}. 
The corrections from the rest terms with stochastic
variable usually contributes up to the fourth order of the coupling
strength $g_{j}$ \cite{Xu2014}. When $g_{j}\ll\omega_{m}$, the
higher order corrections are negligible. Moreover,
it is shown quite clearly in Fig.~\ref{fig:Fi} that the contribution of
$F_{5}$ are always negligible even in the non-Markovian case. 
With the noise-free $O$ operator above (\ref{ofree}), the master equation takes a very simple form as 
\begin{eqnarray}
 &  & \frac{d}{dt}\rho=i\Delta(a^{\dagger}a\rho-\rho a^{\dagger}a)-i\omega_{m}(b^{\dagger}b\rho-\rho b^{\dagger}b)\nonumber \\
 &  & -iG(b^{\dagger}a^{\dagger}\rho-\rho b^{\dagger}a^{\dagger})-iG(b^{\dagger}a\rho-\rho b^{\dagger}a)\nonumber \\
 &  & -iG(ba^{\dagger}\rho-\rho ba^{\dagger})-iG(ba\rho-\rho ba)\nonumber \\
 &  & +\{F_{1}^{*}(b\rho b^{\dagger}-\rho b^{\dagger}b)+F_{2}^{*}(b\rho b-\rho bb)\nonumber \\
 &  & +F_{3}^{*}(b\rho a^{\dagger}-\rho a^{\dagger}b)+F_{4}^{*}(b\rho a-\rho ab)+H.c.\}.\label{eq:MEQ}
\end{eqnarray}
It should be noted that the derivation of the master equation is also irrespective of the
format of the correlation function $\alpha(t,s)$, namely, the master
equation here is applicable to an arbitrary correlation function.
As we have discussed, when $\alpha(t,s)=\delta(t,s)$ (setting $\Gamma=1$), it is straightforward
to show that $F_{1}(t)=0.5$ while $F_{j}(t)=0$ ($j=2,3,4,5$) {[}See
Eq. (\ref{eq:fi}) and (\ref{eq:BC}){]}. Therefore, $O(t,s)=b$,
and equation (\ref{eq:QSD}) is reduced to the traditional Markov
quantum trajectory equation \cite{Dalibardetal,Gisin-Percival}. Correspondingly,
in the master equation (\ref{eq:MEQ}) with $O=b$ is reduced to
\begin{equation}
\frac{d}{dt}\rho=-i[H_{S},\rho]+\{[b,\rho b^{\dagger}]+H.c.\}.
\end{equation}
This is just the standard Lindblad master equation obtained in the
Markov approximation \cite{Markov}.

\section{\label{sec:Numerical-results-and}Numerical results and discussions}

Solving the optomechanical model by the above non-Markovian approaches,
we are capable of analyzing the properties of the entanglement between
the cavity field and movable mirror in a non-Markovian regime. For
a continuous variable system, several separability criteria exist
\cite{CVENT1,CVENT2,CVENT3}. Here, we will employ the logarithmic
negativity \cite{CVENT3} to measure the optomechanical entanglement.
For a two-mode Gaussian state, it is convenient to write down the
momentum operator $p$ and the position operator $q$ in a vector
form as 
\begin{equation}
\xi=(q_{1},p_{1},q_{2},p_{2}),
\end{equation}
where $p_{1}=-i(a-a^{\dagger})$, $q_{1}=(a+a^{\dagger})$, $p_{2}=-i(b-b^{\dagger})$,
$q_{2}=(b+b^{\dagger})$. Then the commutation relations can be written
as 
\begin{equation}
[\xi_{\alpha},\xi_{\beta}]=2iM_{\alpha\beta},
\end{equation}
where 
\begin{equation}
M=\left[\begin{array}{cc}
J & 0\\
0 & J
\end{array}\right],\quad J=\left[\begin{array}{cc}
0 & 1\\
-1 & 0
\end{array}\right].
\end{equation}
The entanglement properties of the two-mode Gaussian state are completely
determined by the variance matrix $V$ which is defined as 
\begin{equation}
V_{\alpha\beta}=\langle\{\Delta\xi_{\alpha},\Delta\xi_{\beta}\}\rangle=\langle(\Delta\xi_{\alpha}\Delta\xi_{\beta}+\Delta\xi_{\beta}\Delta\xi_{\alpha})/2\rangle,
\end{equation}
where $\Delta\xi_{\alpha}=\xi_{\alpha}-\langle\xi_{\alpha}\rangle$.
The variance matrix can be written in a block form as 
\begin{equation}
V=\left[\begin{array}{cc}
A & C\\
C^{T} & B
\end{array}\right].
\end{equation}
Finally, the logarithmic negativity is defined as 
\begin{align}
En(V)= & \max[0,-\ln\nu_{-}],
\end{align}
where $\nu_{-}$ is the smallest eigenvalue of the variance matrix
$V$, which can be computed as 
\begin{equation}
\nu_{-}=\sqrt{[\Sigma(V)-\sqrt{\Sigma(V)^{2}-4\det V}]/2},
\end{equation}
and $\Sigma(V)=\det A+\det B-2\det C$.

In order to compute the logarithmic negativity, we
need to compute a set of mean values of operators by using the non-Markovian
master equation or NMQSD equation, 
\begin{equation}
\frac{d}{dt}\langle A\rangle=\frac{d}{dt}M{[\langle\psi_{t}(z^{*})|A|\psi_{t}(z^{*}))]}={\rm tr}(A\frac{d}{dt}\rho).\label{eq:mean}
\end{equation}
For this particular model, it is more straightforward to use the derived
master equation. However, we pointed out that we can always use the NMQSD equation without
deriving the corresponding master equation. The details of the equations
for the mean values of operators can be found in the Appendixes.

\subsection{Memory enhanced entanglement generation}

\begin{figure}
\noindent \includegraphics[clip,width=1\columnwidth]{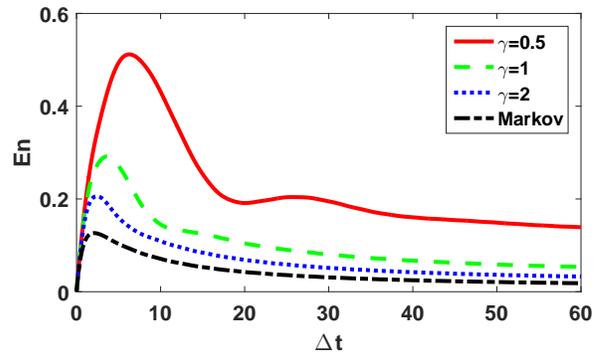}
\caption{\label{fig:NMEntGen}(color online) Memory effect enhanced entanglement
generation. The red (solid), green (dashed), and blue (dotted) lines
are plotted with different memory times $1/\gamma$. The other parameters
are chosen as $\omega_{m}=1$, $\Delta=1$, $G=0.1$, $\Omega=0$,
$\Gamma=2$. The initial state is chosen as $|\psi_{i}\rangle=|00\rangle$.
As a comparison, the Markov case is also plotted as the black (dash-dotted)
line.}
\end{figure}

The memory modulated entanglement dynamics is an interesting problem 
recently \cite{Plenio02,Huelga07,Lambert07,Yi03}.
Therefore, it is desirable to examine how the environmental parameters $\gamma$, $\Omega$
affect the entanglement generation between the optical field and the
mechanical mode. Fig.~\ref{fig:NMEntGen} shows the dynamics of the
entanglement $En$ with different memory times. As a comparison, the
Markov evolution is also plotted in the figure by setting the correlation
function $\alpha(t,s)$ as $\delta(t,s)$. It should be noted that
according to Fig.~\ref{fig:NMEntGen}, we see that a longer memory
time (small $\gamma$) will cause a faster entanglement generation.
Meanwhile, it is found the longer memory times, the longer duration of the optomechanical 
entanglement. Since the major decoherence agent of this model is the
amplitude damping, the environmental memory plays a role of slowing down the
dissipative process due to the back reaction or information back flow.
Therefore, one expects that the dissipative dynamics will experience 
temporal revivals due to the memory effect. On the contrary, the Markov environment
causes the system excitations to decay into the environment exponentially
without any information back flow. More importantly, the non-Markovian
properties of the environment may also affect the residue entanglement
in the steady state ($t\rightarrow \infty$). From Fig.~\ref{fig:NMEntGen}, a longer memory
time give rise to higher residue entanglement degree in a long-time limit.
The Markov steady state entanglement in an optomechanical
system is discussed in many references 
such like \cite{Opto4}.  In the non-Markovian case considered in this paper, 
our results show that the dissipation
and the back flow from the environment may reach a new balance so
that the steady entanglement has a memory of its history. Namely, the steady
entanglement may be dependent on the environmental memory time.
This finding may be understood from the fact the steady states of a non-Markovian
dynamical system are sensitively dependent on the environmental memory 
parameter $\gamma$.  In summary, as seen in the numerical
simulations, the environmental memory can significantly affect the
entanglement generation in both the short-time and long-time limits.

\subsection{Environmental central frequency and entanglement generation}

\begin{figure}[ht]
\noindent \includegraphics[clip,width=1\columnwidth]{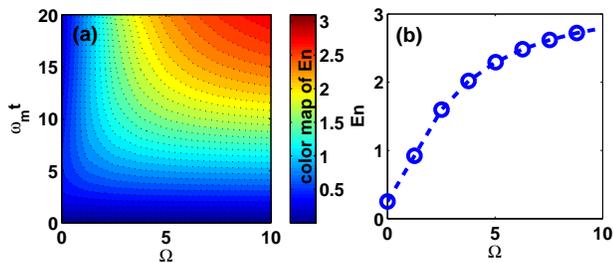} \caption{\label{fig:EntOmega}(color online) Time evolution of entanglement
for different $\Omega$. The left contour plot (a) is the time evolution
of the entanglement indicator $En$ for various values of $\Omega$.
The right 2-D panel (b) is the entanglement generation for different
$\Omega$ at the fixed time $\omega_{m}t=20$. The initial state is
chosen as $|\psi_{i}\rangle=|00\rangle$. The other parameters are
chosen as $\omega_{m}=1$, $\Delta=1$, $G=0.1$, and $\gamma=1$.}
\end{figure}

Apart from the memory time, another important feature of the environment
is dictated by the environmental central frequency $\Omega$, which
is shown to be important to the entanglement generation \cite{Xu2014}.
In Fig.~\ref{fig:EntOmega}, we plot the time evolution of entanglement
for different $\Omega$. The numerical results show that a large
$\Omega$ is useful in generating the optomechanical entanglement. The parameter
$\Omega$ indicates the oscillation frequency of the correlation function $\alpha(t,s)=\frac{\Gamma\gamma}{2}e^{-(\gamma+i\Omega)|t-s|}$,.
A larger $\Omega$ gives rise to a faster oscillation. Therefore, it
explains why a large $\Omega$ can help to preserve the entanglement
since the system is less sensitive to the high-frequency random noise (i.e., when $\Omega$ is large). 
Therefore, the high frequency oscillation effectively causes less entanglement degradation after the
cavity-mirror entanglement is formed. It is worth to note that this
phenomenon can be only observed in non-Markovian case. In the Markov
limit, $\alpha(t,s)=\delta(t,s)$, the $O$ operator becomes a time-independent
function with constant coefficients. This is an important feature showing the remarkable difference
between the non-Markovian and Markov cases. In the non-Markovian
case,  the information backflow from the environment to the system can effective
protect the entanglement, while in the Markov case the dissipation is monotonic,
and the information once dissipated into environment will never come back
to the system of interest.

In an experiment context, a new engineering technique about simulating a non-Markovian
environment shed a new light on controlling the environment memory effect \cite{Liu_2011_NatPhys}.
This new findings are certainly of interest for motivating more theoretical studies on artificial non-Markovian environment
 For example, in the precise quantum measurement \cite{Yang2012},
the probe can be an effective environment with highly non-Markovian
features. In a similar fashion, one can view a pseudomode coupled to an external Markov reservoir
as an effectively non-Markovian environment. 

\subsection{Entanglement generation and the detuning}

\begin{figure}[ht]
\noindent \includegraphics[clip,width=1\columnwidth]{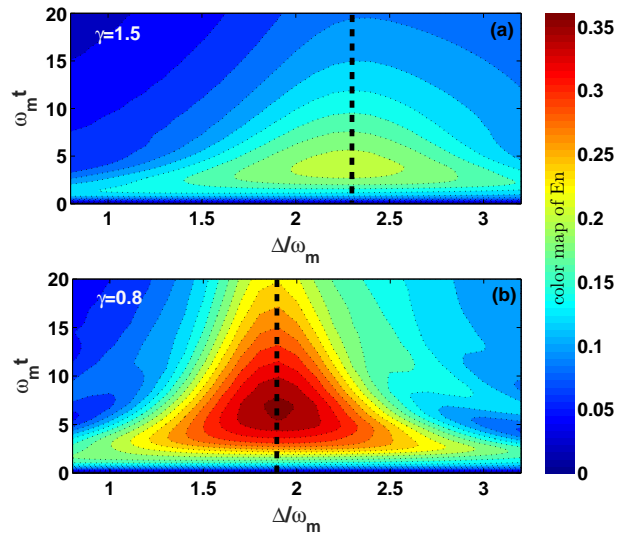} \caption{\label{fig:EntDelta}(color online) Time evolution of entanglement
for different effective detuning $\Delta/\omega_{m}$. The initial
state is chosen as $|\psi_{i}\rangle=|00\rangle$. (a) and (b) are plotted for different memory time $\gamma=1.5$ and $\gamma=0.8$ respectively. The other parameters
are chosen as $\omega_{m}=1$, $\Omega=0$, $G=0.1$, and $\Gamma=4$.}
\end{figure}

In Fig.~\ref{fig:EntDelta}, we illustrate the dynamics of the entanglement as the
function of effective laser detuning $\Delta$ and  $\omega_{m}t$. 
Comparing with the environmental spectrum, the driving laser detuning
is a more convenient parameter that is effectively controllable.
In order to achieve the maximum entanglement generation,
one needs to adjust the effective detuning properly. More importantly,
the choice of effective detuning substantially depends on the non-Markovian
properties of the environment. As illustrated in Fig.~\ref{fig:EntDelta}
(a), when the memory effect parameter $\gamma=1.5$ (relatively weak
non-Markovian case), the maximum entanglement appears at $\Delta/\omega_{m}\approx2.3$.
As a comparison, in Fig.~\ref{fig:EntDelta}~(b), when the memory
effect parameter $\gamma=0.8$ (relatively strong non-Markovian),
the maximum entanglement appears at $\Delta/\omega_{m}\approx1.9$.
The multiple-dependance of the entanglement generation on the parameters $\Delta$ and $\omega$ 
shows that the optimal entanglement generation in an experiment may benefit from the
detailed analysis of the parameter space.  Given the non-Markovian environment, one needs
to choose a suitable laser detuning in order to generate
the maximum entanglement. It was shown in Ref.~\cite{Opto4}, the choice
of effective laser detuning depends on several parameters of the optomechanical
system. Here, we emphasize that the non-Markovian properties 
is also an important factor that can significantly affect the choice
of effective detuning. The results in Fig.~\ref{fig:EntDelta} may
prove to be useful in the future when an experiment on generating optomechanical
entanglement in non-Markovian environment is conducted.

\section{Conclusion and summary}
\label{conclusion}

The recent progress in the experiment
has shown that the optomechanical system can be realized in several
interesting settings including  the traditional cavity-QED systems
as well as some artificial circuit-QED systems \cite{circuitQED}.
More remarkably, with the development of new technology, the environment
can be engineered to purposely control the properties of the desired
quantum entanglement \cite{Liu_2011_NatPhys}. Thus, it is useful 
to develop a versatile theoretical protocols to manipulate entanglement
based on the non-Markovian features of the environment.

Our presented results show that the entanglement of the optomechanical
system can be strongly affected by several features dictated by a non-Markovian environment. 
As we showed in
Fig.~\ref{fig:NMEntGen} and Fig.~\ref{fig:EntOmega}, the entanglement
dynamics of the optomechanical system is sensitively dependent on
the choice of the model parameters. And the optomechanical entanglement 
can be generated in many different ways sensibly depending on the correlation
time and the environmental central frequency as well as the the detuning.
Our analysis is not expected to unveil all interesting aspects of
the environment effect on the optomechanical systems. Rather, our
current research should be regarded as an attempt to incorporate the
environment memory effects in a more systematic way. Indeed, as shown in this
paper, the standard Markov approximation is not adequate for many
interesting physical systems. This is particularly important for the macroscopic
system as the decoherence time could be comparable with the Markov time, so
the Markov approximations deemed to be inadequate when the temporal behaviors
of entanglement is of interest. We show that the residue entanglement for the
non-Markovian cases can be distinctively different from the Markov steady state.

In summary, we have presented a proposal to entangle a macroscopic vibrating
mirror with a cavity field by taking the environment memory effects
into account. We show how to use the NMQSD method to solve the
model involving an optomechanical system coupled to a non-Markovian
environment. In particular, entanglement generation and duration are
fully investigated in non-Markovian regimes.  We conclude
by pointing out a further research is currently being conducted on
the physical models where the quantum system-bath coupling is beyond RWA approximation.

\begin{acknowledgments}
We acknowledge grant support from DOD/AF/AFOSR No. FA9550-12-1-0001.
The research was also supported by the National Natural Science Foundation
of China (Grant No. 11205056 \& No. 11204078) and the Fundamental
Research Funds for the Central Universities (Grant No. 2014ZZD10 \&
No. 2015MS55).
\end{acknowledgments}
\appendix

\section{\label{sec:App1}Optomechanical system with both cavity leakage and
thermal damping of the mirror}

Here, we consider a general case that contains all the possible decoherence
mechanism of the system, namely, the $L$ operator in Eq. (\ref{eq:HI})
can be written as $L=a+b$. In order to incorporate the finite-temperature
bath, we can transform the finite temperature case into an effective
zero temperature model by introducing a fictitious bath \cite{Yu_FiniteT}.
The initial state of the bath for finite temperature case is in the
thermal equilibrium state $\rho_{B}(0)=\frac{e^{-\beta H_{B}}}{Z}$,
where $Z={\rm tr}[e^{-\beta H_{B}}]$ is the partition function with
$\beta=1/k_{B}T$. The occupation number for mode $k$ should be 
\begin{equation}
\langle c_{k}^{\dagger}c_{k}\rangle=\bar{n}_{k}=\frac{1}{e^{-\beta\omega_{k}}-1},\label{eq:BEdis}
\end{equation}
which is the well known Bose-Einstein distribution. By introducing
a fictitious bath $H_{C}=-\sum_{k}\omega_{k}c_{k}^{\prime\dagger}c_{k}^{\prime}$,
without direct interaction with the system and the real bath $H_{B}$,
it is possible to map the finite temperature problem into a zero temperature
problem with two individual baths. Under the Bogoliubov transformation
\begin{equation}
c_{k}=\sqrt{\bar{n}_{k}+1}d_{k}+\sqrt{\bar{n}_{k}}e^{\dagger},
\end{equation}
\begin{equation}
c_{k}^{\prime}=\sqrt{\bar{n}_{k}+1}e_{k}+\sqrt{\bar{n}_{k}}d^{\dagger},
\end{equation}
it is easy to check the vacuum state $|0\rangle=|0\rangle_{d}\otimes|0\rangle_{e}$
satisfies $\langle0|_{d}\langle0|_{e}c_{k}^{\dagger}c_{k}|0\rangle_{d}|0\rangle_{e}=\bar{n}_{k}$.
Therefore, solving the original model plus fictitious bath $H_{tot}=H_{S}+H_{B}+H_{I}+H_{C}$
with the initial vacuum state $|0\rangle_{d}\otimes|0\rangle_{e}$
is equivalent to solving $H_{S}+H_{B}+H_{I}$ with the thermal initial
state $\rho_{B}(0)=e^{-\beta H_{B}}/Z$. Now, we need to solve the
Hamiltonian in the interaction picture as 
\begin{align}
H_{\mathrm{tot}}^{(I)}(t) & =H_{\mathrm{s}}+\sum_{k}(f_{k}e^{-i\omega_{k}t}L^{\dagger}d_{k}+f_{k}e^{i\omega_{k}t}Ld_{k}^{\dagger})\nonumber \\
 & +\sum_{k}(h_{k}e^{-i\omega_{k}t}L^{\dagger}e_{k}^{\dagger}+h_{k}e^{i\omega_{k}t}Le_{k}),\label{eq:FTHtot}
\end{align}
where $f_{k}=\sqrt{\bar{n}_{k}+1}g_{k}$ and $h_{k}=\sqrt{\bar{n}_{k}}g_{k}$
are the effective coupling constants, and the Lindblad operator is
\begin{equation}
L=a+b,
\end{equation}
implying both the cavity leakage and the mirror damping are all taken
into consideration.

The non-Markovian NMQSD equation for the finite-temperature case is
given by \cite{Yu_FiniteT},
\begin{align}
 & \frac{\partial}{\partial t}|\psi(t,z^{*},w^{*})\rangle=\Big[-iH_{\mathrm{s}}+Lz_{t}^{\ast}+L^{\dagger}w_{t}^{\ast}\nonumber \\
 & -L^{\dagger}\int\nolimits _{0}^{t}ds\alpha_{1}(t,s)\frac{\delta}{\delta z_{s}^{\ast}}\nonumber \\
 & -L\int\nolimits _{0}^{t}ds\alpha_{2}(t,s)\frac{\delta}{\delta w_{s}^{\ast}}\Big]|\psi(t,z^{*},w^{*})\rangle,\label{eq:FTQSD}
\end{align}
where $z_{t}^{\ast}=-i\sum_{k}f_{k}z_{k}^{\ast}e^{i\omega_{k}t}$,
$w_{t}^{\ast}=-i\sum_{k}h_{k}^{*}w_{k}^{\ast}e^{-i\omega_{k}t}$ are
two statistically independent Gaussian noises, and $\alpha_{1}(t,s)=\sum_{k}|f_{k}|^{2}e^{-i\omega_{k}(t-s)}$
and $\alpha_{2}(t,s)=\sum_{k}|h_{k}|^{2}e^{i\omega_{k}(t-s)}$ are
correlation functions for the two effective baths. Then, we can replace
the functional derivatives in Eq. (\ref{eq:FTQSD}) by two $O$ operators
\begin{equation}
O_{1}(t,s,z^{\ast},w^{\ast})|\psi(t,z^{*},w^{*})\rangle=\frac{\delta}{\delta z_{s}^{\ast}}|\psi(t,z^{*},w^{*})\rangle,
\end{equation}
\begin{equation}
O_{2}(t,s,z^{\ast},w^{\ast})|\psi(t,z^{*},w^{*})\rangle=\frac{\delta}{\delta w_{s}^{\ast}}|\psi(t,z^{*},w^{*})\rangle,
\end{equation}
and the $O$ operators satisfy the following equations, 
\begin{eqnarray}
\frac{\partial}{\partial t}O_{1} & = & [-iH_{\mathrm{s}}+Lz_{t}^{\ast}+L^{\dagger}w_{t}^{\ast}-L^{\dagger}\bar{O}_{1}-L\bar{O}_{2},O_{1}]\nonumber \\
 &  & -L^{\dagger}\frac{\delta}{\delta z_{s}^{\ast}}\bar{O}_{1}-L\frac{\delta}{\delta z_{s}^{\ast}}\bar{O}_{2},\label{eq:FTO1}\\
\frac{\partial}{\partial t}O_{2} & = & [-iH_{\mathrm{s}}+Lz_{t}^{\ast}+L^{\dagger}w_{t}^{\ast}-L^{\dagger}\bar{O}_{1}-L\bar{O}_{2},O_{2}]\nonumber \\
 &  & -L^{\dagger}\frac{\delta}{\delta w_{s}^{\ast}}\bar{O}_{1}-L\frac{\delta}{\delta w_{s}^{\ast}}\bar{O}_{2},\label{eq:FTO2}
\end{eqnarray}
with the initial condition $O_{1}(t,s=t,z^{\ast},w^{\ast})=L$ and
$O_{2}(t,s=t,z^{\ast},w^{\ast})=L^{\dagger}$, where $\bar{O}_{i}=\int_{0}^{t}\alpha_{i}(t,s)O_{i}(t,s,z^{\ast},w^{\ast})ds$
($i=1,2$). For this particular model, the $O$ operator can be solved
as
\begin{widetext}
\begin{eqnarray}
O_{i} & (t,s,s^{\prime})= & x_{i1}(t,s)a+x_{i2}(t,s)a^{\dagger}+x_{i3}(t,s)b+x_{i4}(t,s)b^{\dagger}\nonumber \\
 &  & +\int_{0}^{t}y_{i1}(t,s,s^{\prime})z_{s^{\prime}}^{*}ds^{\prime}+\int_{0}^{t}y_{i2}(t,s,s^{\prime})w_{s^{\prime}}^{*}ds^{\prime}\;(i=1,2),\label{eq:Ofull}
\end{eqnarray}

\begin{eqnarray}
\frac{\partial}{\partial t}x_{i1}(t,s) & = & -i\Delta x_{i1}+iGx_{i3}-iGx_{i4}+X_{11}x_{i1}-2X_{21}x_{i2}+X_{22}x_{i1}\nonumber \\
 &  & -X_{23}x_{i4}+X_{24}x_{i3}+X_{11}x_{i3}-X_{21}x_{i4}-Y_{2i}(t,s),\label{eq:dxi1}
\end{eqnarray}
\begin{eqnarray}
\frac{\partial}{\partial t}x_{i2}(t,s) & = & i\Delta x_{i2}+iGx_{i3}-iGx_{i4}-X_{11}x_{i2}+2X_{12}x_{i1}-X_{13}x_{i4}\nonumber \\
 &  & +X_{14}x_{i3}-X_{22}x_{i2}+X_{12}x_{i3}-X_{22}x_{i4}-Y_{1i}(t,s),
\end{eqnarray}
\begin{eqnarray}
\frac{\partial}{\partial t}x_{i3}(t,s) & = & i\omega_{m}x_{i3}+iGx_{i1}-iGx_{i2}+X_{13}x_{i1}-X_{23}x_{i2}+X_{13}x_{i3}\nonumber \\
 &  & -X_{21}x_{i2}+X_{22}x_{i1}-2X_{23}x_{i4}+X_{24}x_{i3}-Y_{2i}(t,s),
\end{eqnarray}
\begin{eqnarray}
\frac{\partial}{\partial t}x_{i4}(t,s) & = & -i\omega_{m}x_{i4}+iGx_{i1}-iGx_{i2}+X_{14}x_{i1}-X_{24}x_{i2}-X_{11}x_{i2}\nonumber \\
 &  & +X_{12}x_{i1}-X_{13}x_{i4}+2X_{14}x_{i3}-X_{24}x_{i4}-Y_{1i}(t,s),
\end{eqnarray}
\begin{equation}
\frac{\partial}{\partial t}y_{ik}(t,s,s^{\prime})=Y_{1k}(t,s^{\prime})(x_{i1}-x_{i3})+Y_{2k}(x_{i4}-x_{i2})\;(i,k=1,2)
\end{equation}

\end{widetext}
where $X_{ij}=\int_{0}^{t}\alpha_{i}(t,s)x_{ij}(t,s)ds$, and $Y_{kl}=\int_{0}^{t}\alpha_{k}(t,s)y_{kl}(t,s,s^{\prime})ds$.
The boundary conditions for these equations are:
\begin{equation}
x_{11}(t,t)=x_{13}(t,t)=x_{22}(t,t)=x_{24}(t,t)=1,
\end{equation}
\begin{equation}
x_{12}(t,t)=x_{14}(t,t)=x_{21}(t,t)=x_{23}(t,t)=0,
\end{equation}
\begin{equation}
y_{ij}(t,t,s^{\prime})=0.
\end{equation}
\begin{equation}
y_{i1}(t,s,t)=x_{i2}(t,s)+x_{i4}(t,s)
\end{equation}
\begin{equation}
y_{i2}(t,s,t)=-x_{i1}(t,s)-x_{i3}(t,s)\label{eq:ini_yij}
\end{equation}
Given Eq.~(\ref{eq:Ofull}-\ref{eq:ini_yij}), $O$ operator can be
fully determined, therefore Eq. (\ref{eq:FTQSD}) is solved.

Using Eq. (\ref{eq:FTQSD}) with the exact $O$ operator in Eq. (\ref{eq:Ofull}),
one can also derive a master equation as 
\begin{eqnarray}
\frac{\partial}{\partial t}\rho_{S} & = & -i[H_{S},\rho_{S}]+[L,M\{P_{t}\bar{O}_{1}^{\dagger}\}]-[L^{\dagger},M\{\bar{O}_{1}P_{t}\}]\nonumber \\
 &  & +[L^{\dagger},M\{P_{t}\bar{O}_{2}^{\dagger}\}]-[L,M\{\bar{O}_{2}P_{t}\}],\label{eq:MEQFT}
\end{eqnarray}
where $P_{t}\equiv|\psi(t,z^{*},w^{*})\rangle\langle\psi(t,z,w)|$
is the stochastic density operator. For the details of deriving the
master equation, one can follow the examples in Refs. \cite{QBM,Ncav,chen2014}.

\section{Equations for mean values}

In this section, we show the methods we have used to compute the evolution
of the entanglement. According to Eq.~(\ref{eq:MEQ}) and Eq.~(\ref{eq:mean}), the equations for the mean values of the operators can be obtained as 
\begin{equation}
\frac{d}{dt}\langle b\rangle=-i\omega_{m}\langle b\rangle-iG\langle a^{\dagger}\rangle-iG\langle a\rangle-\sum_{i=1}^{4}F_{i}\langle O_{i}\rangle\label{dq1}
\end{equation}
\begin{equation}
\frac{d}{dt}\langle b^{\dagger}\rangle=i\omega_{m}\langle b^{\dagger}\rangle+iG\langle a^{\dagger}\rangle+iG\langle a\rangle-\sum_{i=1}^{4}F_{i}^{*}\langle O_{i}^{\dagger}\rangle
\end{equation}
\begin{equation}
\frac{d}{dt}\langle a\rangle=i\Delta\langle a\rangle-iG\langle b^{\dagger}\rangle-iG\langle b\rangle
\end{equation}
\begin{equation}
\frac{d}{dt}\langle a^{\dagger}\rangle=-i\Delta\langle a^{\dagger}\rangle+iG\langle b^{\dagger}\rangle+iG\langle b\rangle
\end{equation}
\begin{equation}
\frac{d}{dt}\langle aa\rangle=2i\Delta\langle aa\rangle-2iG\langle ab^{\dagger}\rangle-2iG\langle ab\rangle
\end{equation}
\begin{equation}
\frac{d}{dt}\langle aa^{\dagger}\rangle=iG\langle ab^{\dagger}\rangle+iG\langle ab\rangle-iG\langle a^{\dagger}b^{\dagger}\rangle-iG\langle a^{\dagger}b\rangle
\end{equation}
\begin{eqnarray}
\frac{d}{dt}\langle ab\rangle & = & i\Delta\langle ab\rangle-i\omega_{m}\langle ab\rangle-iG(\langle aa^{\dagger}\rangle+\langle bb^{\dagger}\rangle-1)\nonumber \\
 &  & -iG\langle aa\rangle-iG\langle bb\rangle-\sum_{i=1}^{4}F_{i}\langle aO_{i}\rangle
\end{eqnarray}
\begin{eqnarray}
\frac{d}{dt}\langle ab^{\dagger}\rangle & = & i\Delta\langle ab^{\dagger}\rangle+i\omega_{m}\langle ab^{\dagger}\rangle-iG\langle b^{\dagger}b^{\dagger}\rangle-iG(\langle bb^{\dagger}\rangle\nonumber \\
 &  & -\langle aa^{\dagger}\rangle)+iG\langle aa\rangle-\sum_{i=1}^{4}F_{i}^{*}\langle O_{i}^{\dagger}a\rangle
\end{eqnarray}
\begin{equation}
\frac{d}{dt}\langle a^{\dagger}a^{\dagger}\rangle=-2i\Delta\langle a^{\dagger}a^{\dagger}\rangle+2iG\langle a^{\dagger}b^{\dagger}\rangle+2iG\langle a^{\dagger}b\rangle
\end{equation}
\begin{eqnarray}
\frac{d}{dt}\langle a^{\dagger}b\rangle & = & -i\Delta\langle a^{\dagger}b\rangle-i\omega_{m}\langle a^{\dagger}b\rangle-iG\langle a^{\dagger}a^{\dagger}\rangle-iG(\langle aa^{\dagger}\rangle\nonumber \\
 &  & -\langle bb^{\dagger}\rangle)+iG\langle bb\rangle-\sum_{i=1}^{4}F_{i}\langle a^{\dagger}O_{i}\rangle
\end{eqnarray}
\begin{equation}
\frac{d}{dt}\langle a^{\dagger}b^{\dagger}\rangle=[\frac{d}{dt}\langle ab\rangle]^{\dagger}
\end{equation}
\begin{equation}
\frac{d}{dt}\langle bb\rangle=-2i\omega_{m}\langle bb\rangle-2iG\langle a^{\dagger}b\rangle-2iG\langle ab\rangle-\sum_{i=1}^{4}2F_{i}\langle bO_{i}\rangle
\end{equation}
\begin{eqnarray}
\frac{d}{dt}\langle bb^{\dagger}\rangle & = & -iG\langle a^{\dagger}b^{\dagger}\rangle-iG\langle ab^{\dagger}\rangle+iG\langle a^{\dagger}b\rangle+iG\langle ab\rangle\nonumber \\
 &  & -\{\sum_{i=1}^{4}F_{i}^{*}\langle O_{i}^{\dagger}b\rangle+H.c.\}
\end{eqnarray}
\begin{eqnarray}
\frac{d}{dt}\langle b^{\dagger}b^{\dagger}\rangle & = & 2i\omega_{m}\langle b^{\dagger}b^{\dagger}\rangle+2iG\langle ab^{\dagger}\rangle+2iG\langle a^{\dagger}b^{\dagger}\rangle\nonumber \\
 &  & -\sum_{i=1}^{4}2F_{i}^{*}\langle O_{i}^{\dagger}b^{\dagger}\rangle
\end{eqnarray}
With the equations above, the $V$ matrix as well as the entanglement
can be computed.


\begin{thebibliography}{10}
\bibitem{Cat}E. Schr\"{o}dinger, \textit{Naturwissenschaften} \textbf{23},
807 (1935).

\bibitem{Macro1}S. Pirandola, D. Vitali, P. Tombesi, and S. Lloyd,
\textit{Phys. Rev. Lett.} \textbf{97}, 150403 (2006).

\bibitem{Macro2} F. Plastina, R. Fazio, and G. M. Palma, \textit{Phys.
Rev. B} \textbf{64}, 113306 (2001).

\bibitem{Macro3}W. D\"{u}r and H.-J. Briegel, \textit{Phys. Rev. Lett.}
\textbf{92}, 180403 (2004).

\bibitem{Macro4}H. Krauter\textit{ et al.}, \textit{Phys. Rev. Lett.}
\textbf{107}, 080503 (2011).

\bibitem{Macro5}P. Sekatski, M. Aspelmeyer, and N. Sangouard, \textit{Phys.
Rev. Lett.} \textbf{112}, 080502 (2014).

\bibitem{Macro6}S. Mancini, V. Giovannetti, D. Vitali, and P. Tombesi,
\textit{Phys. Rev. Lett.} \textbf{88}, 120401 (2002).

\bibitem{Macro7}L. Zhou, H. Xiong, and M. S. Zubairy, \textit{Phys.
Rev. A} \textbf{74}, 022321 (2006).

\bibitem{Macro8} X.-Y. Zhao, Y.-H. Ma, and L. Zhou, \textit{Opt.
Commun.} \textbf{282}, 1593 (2009).

\bibitem{Macro9}S. Bose, K. Jacobs, and P. L. Knight, \textit{Phys.
Rev. A} \textbf{59}, 3204 (1999).

\bibitem{Macro10}S. Mancini, V. I. Man'ko, and P. Tombesi, \textit{Phys.
Rev. A} \textbf{55}, 3042 (1997).

\bibitem{Chou2008PRE}C.-H. Chou, T. Yu, and B. L. Hu, \textit{Phys.
Rev. E} \textbf{77}, 011112 (2008).

\bibitem{Chou2008Physica}
C. H. Chou, B. L. Hu, and T. Yu, \textit{Physica A} \textit{387}, 432 (2008).

\bibitem{QtoC}K. C. Schwab, and M. L. Roukes, \textit{Phys. Today}
\textbf{58}, 36 (2005).

\bibitem{QtoC2}W. H. Zurek, \textit{Phys. Today} \textbf{44} (10),
36 (1991).

\bibitem{Experi1}S.~Gr\"{o}blacher, K.~Hammerer, M.~R.~Vanner, and
\mbox{M.~Aspelmeyer}, \textit{Nature} \textbf{460}, 724 (2009).

\bibitem{OptoCool1}J. Teufel \textit{et al.}, \textit{Nature (London)}
\textbf{475}, 359 (2011).

\bibitem{Experi3}T. A. Palomaki, J. D. Teufel, R. W. Simmonds, and
K. W. Lehnert, \textit{Science} \textbf{342}, 710 (2013).

\bibitem{Opto1}S. Mancini, D. Vitali, and P. Tombesi, \textit{Phys.
Rev. Lett.} \textbf{80}, 688 (1998).

\bibitem{Opto2}M. Aspelmeyer, T. J. Kippenberg, and F. Marquardt,
\textit{Rev. Mod. Phys.} \textbf{86}, 1391 (2014).

\bibitem{Opto3}W.~Marshall, C.~Simon, R.~Penrose, and D.~Bouwmeester,
\textit{Phys. Rev. Lett.} \textbf{91}, 130401 (2003).

\bibitem{Opto4}D. Vitali \textit{et al.}, \textit{Phys. Rev. Lett.}
\textbf{98}, 030405 (2007).

\bibitem{ChenJPBreview}Y. Chen, \textit{J. Phys. B} \textbf{46},
104001 (2013).

\bibitem{Opto5}C. Genes, A. Mari, P. Tombesi, and D. Vitali, \textit{Phys.
Rev. A} \textbf{78}, 032316 (2008).

\bibitem{Opto6}A. Nunnenkamp, K. Borkje, and S. M. Girvin, \textit{Phys.
Rev. Lett.} \textbf{107}, 063602 (2011).

\bibitem{Opto7}P. Meystre, \textit{Ann. Phys. (Berlin)} \textbf{525},
215 (2013).

\bibitem{Opto8}K. Zhang, F. Bariani, and P. Meystre, \textit{Phys.
Rev. Lett.} \textbf{112}, 150602 (2014).

\bibitem{Opto9}R. Ghobadi, S. Kumar, B. Pepper, D. Bouwmeester, \mbox{A.
I. Lvovsky}, and C. Simon, \textit{Phys. Rev. Lett.} \textbf{112},
080503 (2014).

\bibitem{Opto10}G. Wang, L. Huang, Y.-C. Lai, and C. Grebogi, \textit{Phys.
Rev. Lett.} \textbf{112}, 110406 (2014).

\bibitem{Markov}H.-P. Breuer and F. Petruccione, \emph{The Theory
of Open Quantum Systems} (Oxford University Press, Oxford, 2002).

\bibitem{Paz2008}J. P. Paz, \textit{Phys. Rev. Lett.} \textbf{100},
220401 (2008).




\bibitem{NMOpto}Y.-D. Wang and A. A. Clerk, \textit{Phys. Rev. Lett.}
\textbf{110}, 253601 (2013).

\bibitem{Groblacher_2015_NMexp}S. Gr\"{o}blacher, A. Trubarov, N.
Prigge, G. D. Cole, M.~Aspelmeyer, and J.~Eisert, \textit{Nat. Commu.} \textbf{6},
7606 (2015).

\bibitem{Cheng_2016_NMOpt}J. Cheng, W. Z. Zhang, L. Zhou, and W.
Zhang, \textit{Sci. Rep.} \textbf{6}, 23678 (2016).

\bibitem{Liu_2011_NatPhys}B.-H. Liu, L. Li, Y.-F. Huang, C.-F. Li,
G.-C. Guo, E.-M. Laine, H.-P. Breuer, and J. Piilo, \textit{Nat. Phys.} \textbf{7},
931 (2011).

\bibitem{Yang2012}H. Yang, H. Miao, and Y. Chen, \textit{Phys. Rev.
A} \textbf{85}, 040101(R) (2012).

\bibitem{QSD98}L. Di\'{o}si, N. Gisin, and W. T. Strunz, \textit{Phys.
Rev. A} \textbf{58}, 1699 (1998).

\bibitem{QSD99}W. T. Strunz, L. Di\'{o}si, and N. Gisin, \textit{Phys.
Rev. Lett.} \textbf{82}, 1801 (1999).

\bibitem{Yu1999}T. Yu, L. Di\'{o}si, N. Gisin, and W. T. Strunz, \textit{Phys.
Rev. A} \textbf{60}, 91 (1999).

\bibitem{QBM}W. T. Strunz and T. Yu, \textit{Phys. Rev. A} \textbf{69},
052115 (2004).

\bibitem{Yu_FiniteT}T. Yu, \textit{Phys. Rev. A} \textbf{69}, 062107
(2004).

\bibitem{Jing-Yu2010}J. Jing and T. Yu, \textit{Phys. Rev. Lett.}
\textbf{105}, 240403 (2010).

\bibitem{Xinyu2011}X. Zhao, J. Jing, B. Corn, and T. Yu, \textit{Phys.
Rev. A} \textbf{84}, 032101 (2011).

\bibitem{ZhaoFB}X. Zhao, W. Shi, L.-A. Wu, and T. Yu, \textit{Phys.
Rev. A} \textbf{86}, 032116 (2012).

\bibitem{ShiFB}W. Shi, X. Zhao, and T. Yu, \textit{Phys. Rev. A}
\textbf{87}, 052127 (2013).

\bibitem{ChenFB}M. Chen, and J. Q. You, \textit{Phys. Rev. A} \textbf{87},
052108 (2013).

\bibitem{Ncav}X. Zhao, J. Jing, J. Q. You, and T. Yu, \textit{Quantum
Inf. and Compu.} \textbf{14}, 0741 (2014).




\bibitem{chen2014}Y. Chen, J. Q. You, and T. Yu, \textit{Phys. Rev.
A} \textbf{90}, 052104 (2014).

\bibitem{Xu2014}J. Xu, X. Zhao, J. Jing, L.-A. Wu, and T. Yu, \textit{J.
Phys. A} \textbf{47}, 435301 (2014).

\bibitem{Law1994}C. K. Law, \textit{Phys. Rev. A} \textbf{49}, 433
(1994).

\bibitem{Joshi2014_RWA}C. Joshi, P. \"{O}hberg, J. D. Cresser, and
E. Andersson, \textit{Phys. Rev. A} \textbf{90}, 063815 (2014).

\bibitem{Cresser1992}J. D. Cresser, \textit{J. Mod. Opt.} \textbf{39},
2187 (1992).

\bibitem{Dalibardetal}J. Dalibard, Y. Castin, and K. M\o{}lmer, \textit{Phys.
Rev. Lett.} \textbf{68}, 580 (1992).

\bibitem{Gisin-Percival}N. Gisin and I.~C. Percival, \textit{J.
Phys. A} \textbf{25}, 5677 (1992).

\bibitem{CVENT1}R. Simon, \textit{Phys. Rev. Lett.} \textbf{84},
2726 (2000).

\bibitem{CVENT2}L.-M. Duan, G. Giedke, J. I. Cirac, and P. Zoller,
\textit{Phys. Rev. Lett.} \textbf{84}, 2722 (2000).

\bibitem{CVENT3}G. Adesso, A. Serafini, and F. Illuminati, \textit{Phys.
Rev. A} \textbf{70}, 022318 (2004).

\bibitem{Naeini_2013_Lorentzian1}A. H. Safavi-Naeini \textit{et al}.,
\textit{New J. Phys.} \textbf{15}, 035007 (2013).

\bibitem{Plenio02}M. B. Plenio and S. F. Huelga, \textit{Phys. Rev.
Lett.} \textbf{88}, 197901 (2002).

\bibitem{Huelga07}S. F. Huelga and M. B. Plenio, \textit{Phys. Rev.
Lett.} \textbf{98}, 170601 (2007).

\bibitem{Lambert07}N. Lambert, R. Aguado, and T. Brandes, \textit{Phys.
Rev. B} \textbf{75}, 045340 (2007).

\bibitem{Yi03}X. X. Yi, C. S. Yu, L. Zhou, and H. S. Song, \textit{Phys.
Rev. A} \textbf{68}, 052304 (2003).

\bibitem{circuitQED}S. Felicetti, \textit{et al}., \textit{Phys.
Rev. Lett.} \textbf{113}, 093602 (2014).\end{thebibliography}
\end{document}